# Quantum private comparison via cavity QED

Tian-Yu Ye*

College of Information & Electronic Engineering, Zhejiang Gongshang University, Hangzhou 310018, P.R.China
*E-mail：happyyty@aliyun.com

**Abstract:** The first quantum private comparison (QPC) protocol via cavity quantum electrodynamics (QED) is proposed in this paper by making full use of the evolution law of atom via cavity QED, where the third party (TP) is allowed to misbehave on his own but cannot conspire with either of the two users. The proposed protocol adopts two-atom product states rather than entangled states as the initial quantum resource, and only needs single-atom measurements for two users. Both the unitary operations and the quantum entanglement swapping operation are not necessary for the proposed protocol. The proposed protocol can compare the equality of one bit from each user in each round comparison with one two-atom product state. The proposed protocol can resist both the outside attack and the participant attack. Particularly, it can prevent TP from knowing two users' secrets. Furthermore, the qubit efficiency of the proposed protocol is as high as 50%.

## 1 Introduction

In the past two decades, quantum information processing had greatly aroused the interest of researchers throughout the world and gained considerable developments. Quantum cryptography, which was first proposed by Bennett and Brassard [1] in 1984, is the development outcome of quantum information processing in the realm of cryptography. It can attain unconditional security on the basis of quantum mechanics principles such as the Heisenberg uncertainty principle, the quantum no-cloning theorem and so on, and has greatly attracted the attentions of many researchers. Up to now, it has established many branches, such as quantum key distribution (QKD) [1-10], quantum secret sharing (QSS) [11-15], quantum secure direct communication (QSDC) [16-24], quantum dialogue (QD) [25-31], quantum teleportation (QT) [32-37] *etc*.

Quantum private comparison (QPC), which was first suggested by Yang *et al.* [38] in 2009, is another interesting branch of quantum cryptography. It aims to compare the equality of secrets from different users without leaking them out on the basis of quantum mechanics principles. As pointed out by Lo [39] that it is impossible to securely evaluate the equality function in a two-party scenario, QPC always requires some additional assumptions such as a third party (TP), *etc*. In recent years, QPC has been greatly developed so that different quantum states and different quantum technologies have been utilized to construct different kinds of QPC protocols. There are QPC protocols with single particles [40-43], Bell states [38,42,44-52], GHZ states [53-56], W states [52,57-58], cluster states [59-60], $\chi$-type entangled states [61-63]. There are QPC protocols requiring the unitary operations [38,40,43,45,53-54,57,59-60,62] and QPC protocols requiring the quantum entanglement swapping operation [44,50-52,55,61].

At present, the physical systems for realizing quantum information processing mainly include cavity quantum electrodynamics (QED), ion trap, nuclear magnetic resonance, quantum dot, superconducting system with Josephson effect, *etc*. [64] Many works, such as entanglement detection [65], Bell-state analysis [66-67], entanglement concentration [68-71], entanglement purification [72], entanglement generation [71,73-74] and quantum repeater [75], have been realized via these physical systems. Cavity QED can guarantee a high level of coherence so that it is an ideal candidate for the experiment of quantum information processing. [64] Cavity QED has become one of the most promising physical systems due to its unique contribution.

Based on the above analysis, following some ideas in the protocol of Ref.[46], we put forward the first QPC protocol via cavity QED by making full use of the evolution law of atom via cavity QED. Different from the standard semi-honest TP first proposed by Chen *et al*. [53], in our protocol, TP is assumed to be a more practical third party. That is, TP is allowed to misbehave on his own but cannot conspire with either of the two users. Similar to the protocol of Ref.[41], our protocol adopts product states rather than entangled states as the initial quantum resource. Unlike the protocols of Refs.[38,40,43,45,53-54,57,59-60,62], our protocol does not need the unitary operations. Unlike the protocols of Refs.[44,50-52,55,61], our protocol does not need the quantum entanglement swapping operation.

The rest of this paper is organized as follows. Sec.2 introduces the model of cavity QED. Sec.3 illustrates the QPC protocol via cavity QED. Sec.4 analyzes its correctness, security and qubit efficiency. Sec.5 discusses its comparison with some previous QPC protocols. Sec.6 gives the conclusion.

## 2 Model of cavity QED

Driven by a classical field, two identical two-level atoms can simultaneously interact with a single-mode cavity. The interaction Hamiltonian between the single-mode cavity and the atoms under the rotating-wave approximation can be depicted as [73-74,76-77]

$$H = \omega_0 S_z + \omega_a a^\dagger a + \sum_{j=1}^{2}[g(a^\dagger S_j^- + a S_j^\dagger) + \Omega(S_j^\dagger e^{-iwt} + S_j^- e^{iwt})]. \quad (1)$$

Here, $S_z = (1/2)\sum_{j=1}^{2}(|e_j\rangle\langle e_j| - |g_j\rangle\langle g_j|)$, $S_j^- = |g_j\rangle\langle e_j|$ and $S_j^\dagger = |e_j\rangle\langle g_j|$, where $|g_j\rangle$ and $|e_j\rangle$ are the ground and excited states of the $j$ th atom, respectively. $g$ is the atom-cavity coupling strength. $a$ and $a^\dagger$ are the annihilation and creation operators for the cavity mode, respectively. $\omega_0$, $\omega_a$, $\omega$ and $\Omega$ are the atomic transition frequency, the cavity frequency, the classical field frequency and the Rabi frequency, respectively. $t$ is the interaction time. Given that $\omega = \omega_0$, the evolution operator of the system in the interaction picture can be expressed as [73-74,76-77]

$$U(t) = e^{-iH_0 t} e^{-iH_e t}, \quad (2)$$

where $H_0 = \Omega\sum_{j=1}^{2}(S_j^\dagger + S_j^-)$, and $H_e$ is the effective Hamiltonian. Considering the large detuning case $\delta \gg g$ ($\delta$ is the detuning between $\omega_0$ and $\omega_a$) and the strong driving regime $\Omega \gg \delta, g$, there is no energy exchange between the atomic system and the cavity. As a result, the effects of cavity decay and thermal field are erased. Consequently, the effective interaction Hamiltonian $H_e$ in the interaction picture can be described as [73-74,76-77]

$$H_e = (\lambda/2)\left[\sum_{j=1}^{2}(|e_j\rangle\langle e_j| + |g_j\rangle\langle g_j|) + \sum_{i,j=1, i\neq j}^{2}(S_i^\dagger S_j^- + S_i^\dagger S_j^\dagger + H.C.)\right], \quad (3)$$

where $\lambda = g^2/2\delta$. When two atoms are sent into the above cavity simultaneously, driven by a classical field, they interact with it. If the interaction time and the Rabi frequency are made to satisfy $\lambda t = \pi/4$ and $\Omega t = \pi$, two atoms will have the following evolution:

$$|gg\rangle_{jk} \to \frac{\sqrt{2}}{2} e^{-i\pi/4}(|gg\rangle_{jk} - i|ee\rangle_{jk}), \quad (4)$$

$$|ge\rangle_{jk} \to \frac{\sqrt{2}}{2} e^{-i\pi/4}(|ge\rangle_{jk} - i|eg\rangle_{jk}), \quad (5)$$

$$|eg\rangle_{jk} \to \frac{\sqrt{2}}{2} e^{-i\pi/4}(|eg\rangle_{jk} - i|ge\rangle_{jk}), \quad (6)$$

$$|ee\rangle_{jk} \to \frac{\sqrt{2}}{2} e^{-i\pi/4}(|ee\rangle_{jk} - i|gg\rangle_{jk}). \quad (7)$$

### 3 QPC protocol via cavity QED

Two users, Alice and Bob, want to know whether their secrets are equal or not under the help of TP, who is allowed to misbehave on his own but cannot conspire with either of them. Assume that $X$ and $Y$ are Alice and Bob's secrets, respectively, where $X = \sum_{j=0}^{L-1} x_j 2^j$, $Y = \sum_{j=0}^{L-1} y_j 2^j$, $x_j, y_j \in \{0,1\}$.

Alice and Bob share two common key sequences $K_A$ and $K_B$ individual with length $L$ through the QKD protocols [1-10] beforehand. Here, $K_A^i, K_B^i \in \{0,1\}$, where $K_A^i$ is the $i$ th bit of $K_A$, $K_B^i$ is the $i$ th bit of $K_B$, and $i = 1, 2, \ldots, L$.

The QPC protocol via cavity QED is constructed as follows:

**Step 1:** Alice (Bob) divides the binary representation of $X$ ($Y$) into $L$ groups $G_A^1, G_A^2, \ldots, G_A^L$ ($G_B^1, G_B^2, \ldots, G_B^L$), where each group contains one binary bit.

**Step 2:** TP prepares a quantum state sequence composed of $L$ product states, each of which is randomly in one of the four states $\{|gg\rangle, |ge\rangle, |eg\rangle, |ee\rangle\}$. This quantum state sequence is denoted as $S = [S_a^1 S_b^1, S_a^2 S_b^2, \ldots, S_a^L S_b^L]$. Here, the subscripts $a, b$ represent two atoms in one product state, while the superscripts $1, 2, \ldots, L$ denote the orders of product states in $S$.

**Step 3:** TP sends $S_a^i S_b^i$ ($i = 1, 2, \ldots, L$) into the single-mode cavity described above. Driven by a classical field, the two atoms $S_a^i$ and $S_b^i$ simultaneously interact with the single-mode cavity. TP chooses the Rabi frequency and the interaction time to satisfy $\Omega t = \pi$ and $\lambda t = \pi/4$. As a result, $S_a^i S_b^i$ will undergo the evolution as shown in formulas (4-7). Obviously, $S_a^i$ and $S_b^i$ become entangled together after the evolution. After they fly out the single-mode cavity, TP picks up $S_a^i$ and $S_b^i$ to form sequences $S_a$ and $S_b$, respectively. That is, $S_a = \left[ S_a^1, S_a^2, \ldots, S_a^L \right]$ and $S_b = \left[ S_b^1, S_b^2, \ldots, S_b^L \right]$.

For security check, TP prepares two sets of sample single atoms $D_a$ and $D_b$ randomly in one of the four states $\{|g\rangle, |e\rangle, |+\rangle, |-\rangle\}$. Here, $|\pm\rangle = \frac{1}{\sqrt{2}}(|g\rangle \pm |e\rangle)$. Note that $\{|g\rangle, |e\rangle\}$ is denoted as $Z$ basis while $\{|+\rangle, |-\rangle\}$ is denoted as $X$ basis. Then, TP randomly inserts $D_a$ ($D_b$) into $S_a$ ($S_b$) to form a new sequence $S_a^{'}$ ($S_b^{'}$). Finally, TP sends $S_a^{'}$ ($S_b^{'}$) to Alice (Bob).

After confirming that Alice (Bob) has received $S_a^{'}$ ($S_b^{'}$), TP tells Alice (Bob) the positions and the preparation basis of sample single atoms from $D_a$ ($D_b$). Then, Alice (Bob) measures the sample single atoms in $S_a^{'}$ ($S_b^{'}$) with the basis TP told and tells TP her (his) measurement results. By comparing the initial states of sample single atoms in $S_a^{'}$ ($S_b^{'}$) with Alice's (Bob's) measurement results, TP can judge whether the quantum channel is secure or not during the transmission of $S_a^{'}$ ($S_b^{'}$). If the error rate is unreasonably high, the communication is stopped; otherwise, Alice (Bob) drops out the sample single atoms in $S_a^{'}$ ($S_b^{'}$) to recover $S_a$ ($S_b$), and goes to the next step.

**Step 4:** For the $i$ th ($i$ from 1 to $L$) round comparison, Alice (Bob) measures atom $S_a^i$ ($S_b^i$) with $Z$ basis. The measurement result of $S_a^i$ ($S_b^i$) is coded with one classical bit which is represented by $M_A^i$ ($M_B^i$). Concretely speaking, if the measurement result of $S_a^i$ ($S_b^i$) is $|g\rangle$, $M_A^i$ ($M_B^i$) will be 0; otherwise, if the measurement result of $S_a^i$ ($S_b^i$) is $|e\rangle$, $M_A^i$ ($M_B^i$) will be 1. Then, Alice (Bob) computes $R_A^i = G_A^i \oplus M_A^i \oplus K_A^i$ ($R_B^i = G_B^i \oplus M_B^i \oplus K_B^i$). Here, $\oplus$ is the module 2 operation. Finally, Alice (Bob) tells TP $R_A^i$ ($R_B^i$) publicly.

**Step 5:** TP transforms $S_a^i S_b^i$ prepared in Step 2 into one classical bit $M_T^i$ according to its initial state. Concretely speaking, if the initial state of $S_a^i S_b^i$ is $|gg\rangle$ or $|ee\rangle$, $M_T^i$ will be 0; otherwise, if the initial state of $S_a^i S_b^i$ is $|ge\rangle$ or $|eg\rangle$, $M_T^i$ will be 1. That is, $M_T^i$ represents the parity of $S_a^i$ and $S_b^i$ prepared by TP. Then, TP computes $R^i = R_A^i \oplus R_B^i \oplus M_T^i$, and tells $R^i$ to Alice and Bob publicly.

**Step 6:** After receiving $R^i$, both Alice and Bob compute $R^{i'} = R^i \oplus K_A^i \oplus K_B^i$. They terminate the protocol and conclude that $X \neq Y$ as long as they find out that $R^{i'} \neq 0$; otherwise, they set $i = i+1$ and repeat the protocol from step 4. If they find out that $R^{i'} = 0$ for all $i$ in the end, they will conclude that $X = Y$.

Now it concludes the description of our protocol. For clarity, we further give out the flow chart of our protocol in Fig.1, taking $S_a^i S_b^i$ for example.

It should be emphasized that in this kind of dissipative system, the effects of cavity decay and thermal field are of great concerns. In order to erase the effects of cavity decay and thermal field, we need use the large detuning case and the strong driving regime. That is, $\delta$ should be chosen to satisfy $\delta \gg g$, and $\Omega$ should be chosen to satisfy $\Omega \gg \delta, g$. In this way, there is no energy exchange between the atomic system and the cavity. In addition, in order to make our protocol work well, formulas (4-7) should be established first. Therefore, both of $\lambda t = \pi/4$ and $\Omega t = \pi$ should be also satisfied.

## 4 Analysis

In this section, we analyze our protocol on three aspects including the correctness of its output, the security and the qubit efficiency.

### 4.1 Correctness

In our protocol, under the help of TP, Alice and Bob compare the equality of $G_A^i$ and $G_B^i$ in the $i$ th round comparison by using the evolution law of $S_a^i S_b^i$ via cavity QED. Here, $G_A^i$ is the $i$ th bit of Alice's secret $X$, $G_B^i$ is the $i$ th bit of Bob's secret $Y$, and $S_a^i S_b^i$ is the $i$ th product state in $S$ prepared by TP. For convenience, we use $IS^i$ to denote the initial state of $S_a^i S_b^i$ prepared by TP. Accordingly, $M_T^i$ is the one-bit classical code of $IS^i$. According to the description of our protocol, it is apparent that $M_A^i$ is the one-bit classical code of the $Z$ basis measurement result of $S_a^i$ after evolution in cavity QED, and $M_B^i$ is the one-bit classical code of the $Z$ basis measurement result of $S_b^i$ after evolution in cavity QED. Moreover, it follows that $R_A^i = G_A^i \oplus M_A^i \oplus K_A^i$, $R_B^i = G_B^i \oplus M_B^i \oplus K_B^i$, $R^i = R_A^i \oplus R_B^i \oplus M_T^i$ and $R^{i'} = R^i \oplus K_A^i \oplus K_B^i$, where $K_A^i$ is the $i$ th bit of $K_A$, and $K_B^i$ is the $i$ th bit of $K_B$. The relationships of these essential variables in our protocol are further shown in Table 1. According to formulas (4-7) or Table 1, it is easy to get that $M_A^i \oplus M_B^i \oplus M_T^i = 0$. Consequently, we obtain

$$\begin{aligned}
R^{i'} &= R^i \oplus K_A^i \oplus K_B^i \\
&= \left(R_A^i \oplus R_B^i \oplus M_T^i\right) \oplus K_A^i \oplus K_B^i \\
&= \left(\left(G_A^i \oplus M_A^i \oplus K_A^i\right) \oplus \left(G_B^i \oplus M_B^i \oplus K_B^i\right) \oplus M_T^i\right) \oplus K_A^i \oplus K_B^i \\
&= \left(G_A^i \oplus G_B^i\right) \oplus \left(M_A^i \oplus M_B^i \oplus M_T^i\right) \\
&= G_A^i \oplus G_B^i .
\end{aligned} \quad (8)$$

According to formula (8), if $R^{i'} = 0$, we will get that $G_A^i = G_B^i$; otherwise, we will get that $G_A^i \neq G_B^i$. It can be concluded now that the correctness of the output of our protocol is validated.

Note that similar to the QPC protocol of Ref.[46], our protocol utilizes the parity of two atoms to accomplish the equality comparison of private secrets. Concretely speaking, in the $i$ th round comparison, $G_A^i$ is encrypted with $M_A^i$ and $K_A^i$ to derive $R_A^i$, and $G_B^i$ is encrypted with $M_B^i$ and $K_B^i$ to derive $R_B^i$. Here, $M_A^i$, $K_A^i$, $M_B^i$ and $K_B^i$ play the role of one-time-pad key. After doing the calculations of $R^i = R_A^i \oplus R_B^i \oplus M_T^i$ and $R^{i'} = R^i \oplus K_A^i \oplus K_B^i$, we get the XOR value of $G_A^i$ and $G_B^i$, as shown in formula (8). As a result, the equality comparison result of $G_A^i$ and $G_B^i$ can be easily obtained. It is obvious that formula (8) relies deeply on the XOR relation of $M_A^i$, $M_B^i$ and $M_T^i$, i.e., $M_A^i \oplus M_B^i \oplus M_T^i = 0$. Actually, $M_A^i \oplus M_B^i$ essentially implies the parity of two atoms $S_a^i$ and $S_b^i$ after evolution via cavity QED, while $M_T^i$ represents the parity of $S_a^i$ and $S_b^i$ before evolution via cavity QED. According to formulas (4-7), the parity of $S_a^i$ and $S_b^i$ keeps unchanged during the evolution via cavity QED so that $M_A^i \oplus M_B^i \oplus M_T^i = 0$ is easily established, which guarantees the correctness of the equality comparison result of $G_A^i$ and $G_B^i$. It can be concluded now that the parity of two atoms $S_a^i$ and $S_b^i$ is utilized to accomplish the equality comparison of $G_A^i$ and $G_B^i$.

Apparently, in the whole equality comparison process of $G_A^i$ and $G_B^i$, our protocol employs neither the unitary operations nor the quantum entanglement swapping operation.

### 4.2 Security
The security towards both the outside attack and the participant attack is discussed in this section.

#### 4.2.1 Outside attack
We analyze the outside attack according to each step of our protocol.

As for steps 1, 2 and 6, an outside eavesdropper has no chance to launch an attack, since there is not any quantum transmission or classical transmission in these steps.

As for step 3, there are quantum transmissions as TP sends $S_a^{'}$ ($S_b^{'}$) to Alice (Bob). An outside eavesdropper may try to obtain the two users' secrets by performing the intercept-resend attack, the measure-resend attack, the entangle-measure attack, *etc*. Fortunately, the sample single atoms randomly in one of the four states $\{|g\rangle, |e\rangle, |+\rangle, |-\rangle\}$ are used to detect the outside attack. This security check method is equivalent to the decoy photon technology [78-79], which has been validated in detail in Refs.[28-29]. Note that the decoy photon technology can be regarded as a variant of the BB84 security check method [1], which has also been validated in Ref.[80]. Therefore, the attacks from an outside eavesdropper are invalid in this step.

As for step 4, Alice (Bob) tells TP $R_A^i$ ($R_B^i$) publicly. Apparently, $G_A^i$ ($G_B^i$) is encrypted with $K_A^i$ ($K_B^i$) in this step. However, an outside eavesdropper has no knowledge about $K_A^i$ ($K_B^i$). In this case, she still has no access to $G_A^i$ ($G_B^i$), even though she hears $R_A^i$ ($R_B^i$) from Alice (Bob).

As for step 5, TP tells $R^i$ to Alice and Bob publicly. It can be derived that

$$\begin{aligned}
R^i &= R_A^i \oplus R_B^i \oplus M_T^i \\
&= \left(G_A^i \oplus M_A^i \oplus K_A^i\right) \oplus \left(G_B^i \oplus M_B^i \oplus K_B^i\right) \oplus M_T^i \\
&= \left(G_A^i \oplus K_A^i\right) \oplus \left(G_B^i \oplus K_B^i\right).
\end{aligned} \quad (9)$$

Even though she hears $R^i$ from TP, an outside eavesdropper still cannot extract anything useful about $G_A^i$ or $G_B^i$ in this step, as she has no access to $K_A^i$ and $K_B^i$.

It can be concluded now that our protocol has a high level of security towards the outside attack.

#### 4.2.2 Participant attack

Participant attack is a kind of attack from a dishonest participant. As suggested by Gao et al. [81], participant attack is generally more powerful than outside attack so that it should be paid more attention to. Two cases of participant attack including the attack from a dishonest user and the attack from TP are analyzed here.

**Case 1: the attack from a dishonest user**

In our protocol, Alice' role is similar to Bob's. Without loss of generality, Alice is assumed to be the dishonest user who wants to obtain the secret of the other user.

In step 3, Alice may want to intercept $S_b'$ TP sends to Bob. However, as she has no knowledge about the positions and the preparation basis of sample single atoms from $D_b$, she will inevitably leave her trace and be caught as an outside attacker if she launches this attack.

In step 4, Alice may hear $R_B^i$ from Bob. $S_a^i$ and $S_b^i$ become entangled together after the evolution via cavity QED in step 3. Since Alice does not know the initial product state of $S_a^i S_b^i$ prepared by TP in step 2, she has no knowledge about the entangled state composed by $S_a^i$ and $S_b^i$ after the evolution. According to formulas (4-7), she cannot correctly deduce $M_B^i$ from $M_A^i$. The only thing she can do is to randomly guess the true value of $M_B^i$. Because $G_B^i$ is encrypted with $M_B^i$, she cannot know $G_B^i$ at all.

In step 5, Alice hears $R^i$ from TP. According to formula (8), she can obtain $G_B^i$ by computing $R^{i'} \oplus G_A^i$. However, our protocol has to be terminated as long as Alice and Bob find out that $R^{i'} \neq 0$ for certain $i$. Therefore, Alice can know all the $G_B^i$ where $R^{i'} = 0$ has been derived and only obtain at most 1 bit of the $G_B^i$ where $R^{i'} \neq 0$ has been derived.

**Case 2: the attack from TP**

In our protocol, TP is allowed to try his best to obtain the secrets from two users without conspiring with either of Alice and Bob.

In step 4, Alice (Bob) tells TP $R_A^i$ ($R_B^i$) publicly. Since TP has no access to $K_A^i$ ($K_B^i$), he still cannot extract $G_A^i$ ($G_B^i$) from $R_A^i$ ($R_B^i$).

In step 6, both Alice and Bob do not tell TP $R^{i'}$. However, once the protocol is terminated by Alice and Bob in the middle, TP naturally knows that $X \neq Y$. This case happens with the probability of $1-\left(\frac{1}{2}\right)^{L-1}$. Or if the protocol is not terminated by Alice and Bob until $G_A^L$ and $G_B^L$ are compared, TP has to randomly guess the comparison result between $X$ and $Y$. This case happens with the probability of $\left(\frac{1}{2}\right)^{L-1}$. It can be concluded that TP knows the exact comparison result between $X$ and $Y$ (i.e., $X \neq Y$) with the probability of $1-\left(\frac{1}{2}\right)^{L-1}$.

### 4.3 Qubit efficiency

The qubit efficiency $\eta_e$ here is defined as $\eta_e = \frac{n_c}{n_q}$, where $n_c$ and $n_q$ are the number of compared classical bits and the number of consumed qubits in each round comparison, respectively [14-15]. In our protocol, one two-atom product state can be used to compare one classical bit from each user. Therefore, the qubit efficiency of our protocol is 50%.

## 5 Discussion

We discuss the comparison between our protocol and some previous QPC protocols [38,41,46,52-53,58,60] in this section. The comparison result is summarized in Table 2. Here, the symbol $n$ is the comparison times on the condition that Alice and Bob's secrets are identical.

From Table 2, we can conclude that the advantage of our protocol lies in having the following features simultaneously:

(1) It employs two-particle product states as the initial quantum resource, which are much easier to prepare than entangled states;

(2) It only needs to perform single-atom measurements, which are easier to accomplish than Bell state measurements;

(3) It does not require the unitary operations or the quantum entanglement swapping operation;

(4) It can be automatically halted once Alice and Bob discover the inequality of their secrets in the middle, which means that not all the secrets are necessary to be compared when their secrets are not identical;

(5) It has a high qubit efficiency equal to 50%.

It should be emphasized that our protocol needs the use of QKD method to guarantee the security, just as analyzed in Sec.4.2. Likewise, the QPC protocol of Ref.[38] uses the hash function to guarantee the security. In fact, the QPC protocol in Ref.[53] is not secure, since Alice and Bob can perform the intercept-resend attack to obtain each other's secret without being discovered, just as pointed out by Ref.[54]. The QPC protocol in Ref.[46] is also not secure when TP is a more practical third party who tries his best to extract useful information about two users' secrets with active attacks, just as pointed out by Refs.[47-49]. When TP is a more practical third party than the one assumed, the QPC protocols in Refs.[53,60] have the similar security loophole induced by TP to that in Ref.[46] indicated above. Actually, in this case, the QKD method can be utilized to guarantee the security of the QPC protocols in Refs.[46,53,60].

In addition, different from the QPC protocols in Refs.[38,41,46,52-53,58,60], our protocol is realized via cavity QED, thus it can make full use of the evolution law of atom via cavity QED. Actually, our protocol is the first QPC protocol via cavity QED.

## 6  Conclusion

In this paper, by making full use of the evolution law of atom via cavity QED, we propose the first QPC protocol via cavity QED. Here, TP is allowed to misbehave on his own but cannot conspire with either of the two users. Our protocol adopts two-atom product states rather than entangled states as the initial quantum resource, and only needs single-atom measurements for two users. It needs neither the unitary operations nor the quantum entanglement swapping operation. It can compare the equality of one bit from each user in each round comparison with one two-atom product state. It can resist both the outside attack and the participant attack. Particularly, it can prevent TP from knowing two users' secrets. Moreover, its qubit efficiency is as high as 50%.

**Acknowledgements**

The author would like to thank the anonymous reviewers for their valuable comments that help enhancing the quality of this paper. Funding by the National Natural Science Foundation of China (Grant No.61402407) is gratefully acknowledged.

**Compliance with ethical standards**

Conflict of interest: The authors declare that they have no conflict of interest.

**Appendix:**

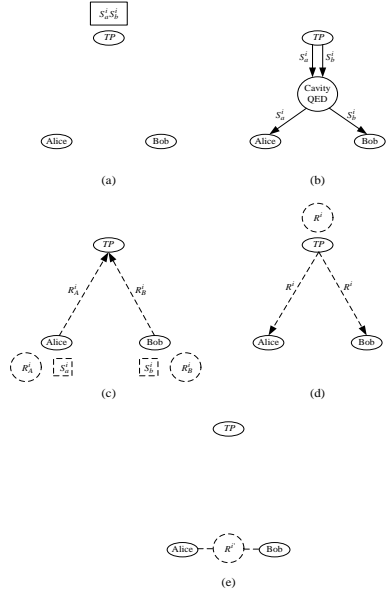

Fig.1  The flow chart of the proposed QPC protocol, taking $S_a^i S_b^i$ for example

(a) TP prepares quantum product state $S_a^i S_b^i$ as the quantum carrier. Here, the solid ellipse denotes an participant, while the solid rectangle denotes the quantum state preparation operation; (b)TP sends atom $S_a^i$ ( $S_b^i$ ) into cavity QED for evolution, and transmits it to Alice (Bob) after it flies out cavity QED. Here, the solid circle denotes cavity QED, while the solid line with an arrow denotes the quantum state transmission operation; (c)Alice (Bob) measures atom $S_a^i$ ( $S_b^i$ ) with Z basis, computes $R_A^i$ ( $R_B^i$ ) and sends $R_A^i$ ( $R_B^i$ ) to TP. Here, the dotted rectangle, the dotted circle and the dotted line with an arrow denote the quantum measurement operation with Z basis, the classical computation operation and the classical information transmission operation, respectively; (d) TP computes $R^i$, and sends $R^i$ to Alice and Bob; (e)Alice and Bob compute $R^{i'}$.

Table 1  The relationships of essential variables in the proposed QPC protocol

( $G_A^i$ is the $i$ th bit of Alice's secret $X$, and $G_B^i$ is the $i$ th bit of Bob's secret $Y$; $IS^i$ denotes the initial state of $S_a^i S_b^i$ prepared by TP, and $M_T^i$ is the one-bit classical code of $IS^i$; $S_a^i S_b^i$ after evolution is the state of $IS^i$ after evolution in cavity QED; $M_A^i$ is the one-bit classical code of the Z basis measurement result of $S_a^i$ after evolution in cavity QED, and $M_B^i$ is the one-bit classical code of the Z basis measurement result of $S_b^i$ after evolution in cavity QED; $K_A^i$ is the $i$ th bit of $K_A$, and $K_B^i$ is the $i$ th bit of $K_B$; $R_A^i = G_A^i \oplus M_A^i \oplus K_A^i$, $R_B^i = G_B^i \oplus M_B^i \oplus K_B^i$, $R^i = R_A^i \oplus R_B^i \oplus M_T^i$ and $R^{i'} = R^i \oplus K_A^i \oplus K_B^i$.)

| $G_A^i$ | $G_B^i$ | $IS^i$ | $M_T^i$ | $S_a^i S_b^i$ (after evolution) | $M_A^i$ | $M_B^i$ | $K_A^i, K_B^i$ | $R_A^i$ | $R_B^i$ | $R^i$ | $R^{i'}$ |
|---|---|---|---|---|---|---|---|---|---|---|---|
| 0 | 0 | $\|gg\rangle$ | 0 | $\|gg\rangle$ | 0 | 0 | 0,0 | 0 | 0 | 0 | 0 |
|   |   |   |   |   |   |   | 0,1 | 0 | 1 | 1 | 0 |
|   |   |   |   |   |   |   | 1,0 | 1 | 0 | 1 | 0 |
|   |   |   |   |   |   |   | 1,1 | 1 | 1 | 0 | 0 |
|   |   |   |   | $\|ee\rangle$ | 1 | 1 | 0,0 | 1 | 1 | 0 | 0 |
|   |   |   |   |   |   |   | 0,1 | 1 | 0 | 1 | 0 |
|   |   |   |   |   |   |   | 1,0 | 0 | 1 | 1 | 0 |
|   |   |   |   |   |   |   | 1,1 | 0 | 0 | 0 | 0 |
|   |   | $\|ge\rangle$ | 1 | $\|ge\rangle$ | 0 | 1 | 0,0 | 0 | 1 | 0 | 0 |
|   |   |   |   |   |   |   | 0,1 | 0 | 0 | 1 | 0 |
|   |   |   |   |   |   |   | 1,0 | 1 | 1 | 1 | 0 |
|   |   |   |   |   |   |   | 1,1 | 1 | 0 | 0 | 0 |
|   |   |   |   | $\|eg\rangle$ | 1 | 0 | 0,0 | 1 | 0 | 0 | 0 |
|   |   |   |   |   |   |   | 0,1 | 1 | 1 | 1 | 0 |
|   |   |   |   |   |   |   | 1,0 | 0 | 0 | 1 | 0 |
|   |   |   |   |   |   |   | 1,1 | 0 | 1 | 0 | 0 |
|   |   | $\|eg\rangle$ | 1 | $\|ge\rangle$ | 0 | 1 | 0,0 | 0 | 1 | 0 | 0 |
|   |   |   |   |   |   |   | 0,1 | 0 | 0 | 1 | 0 |
|   |   |   |   |   |   |   | 1,0 | 1 | 1 | 1 | 0 |
|   |   |   |   |   |   |   | 1,1 | 1 | 0 | 0 | 0 |
|   |   |   |   | $\|eg\rangle$ | 1 | 0 | 0,0 | 1 | 0 | 0 | 0 |
|   |   |   |   |   |   |   | 0,1 | 1 | 1 | 1 | 0 |
|   |   |   |   |   |   |   | 1,0 | 0 | 0 | 1 | 0 |
|   |   |   |   |   |   |   | 1,1 | 0 | 1 | 0 | 0 |

| | | | | | | | | | | | | |
|---|---|---|---|---|---|---|---|---|---|---|---|---|
| | | $\|ee\rangle$ | 0 | $\|gg\rangle$ | 0 | 0 | 0,0 | 0 | 0 | 0 | 0 | |
| | | | | | | | 0,1 | 0 | 1 | 1 | 0 | |
| | | | | | | | 1,0 | 1 | 0 | 1 | 0 | |
| | | | | | | | 1,1 | 1 | 1 | 0 | 0 | |
| | | | | $\|ee\rangle$ | 1 | 1 | 0,0 | 1 | 1 | 0 | 0 | |
| | | | | | | | 0,1 | 1 | 0 | 1 | 0 | |
| | | | | | | | 1,0 | 0 | 1 | 1 | 0 | |
| | | | | | | | 1,1 | 0 | 0 | 0 | 0 | |
| 0 | 1 | $\|gg\rangle$ | 0 | $\|gg\rangle$ | 0 | 0 | 0,0 | 0 | 1 | 1 | 1 | |
| | | | | | | | 0,1 | 0 | 0 | 0 | 1 | |
| | | | | | | | 1,0 | 1 | 1 | 0 | 1 | |
| | | | | | | | 1,1 | 1 | 0 | 1 | 1 | |
| | | | | $\|ee\rangle$ | 1 | 1 | 0,0 | 1 | 0 | 1 | 1 | |
| | | | | | | | 0,1 | 1 | 1 | 0 | 1 | |
| | | | | | | | 1,0 | 0 | 0 | 0 | 1 | |
| | | | | | | | 1,1 | 0 | 1 | 1 | 1 | |
| | | $\|ge\rangle$ | 1 | $\|ge\rangle$ | 0 | 1 | 0,0 | 0 | 0 | 1 | 1 | |
| | | | | | | | 0,1 | 0 | 1 | 0 | 1 | |
| | | | | | | | 1,0 | 1 | 0 | 0 | 1 | |
| | | | | | | | 1,1 | 1 | 1 | 1 | 1 | |
| | | | | $\|eg\rangle$ | 1 | 0 | 0,0 | 1 | 1 | 1 | 1 | |
| | | | | | | | 0,1 | 1 | 0 | 0 | 1 | |
| | | | | | | | 1,0 | 0 | 1 | 0 | 1 | |
| | | | | | | | 1,1 | 0 | 0 | 1 | 1 | |
| | | $\|eg\rangle$ | 1 | $\|ge\rangle$ | 0 | 1 | 0,0 | 0 | 0 | 1 | 1 | |
| | | | | | | | 0,1 | 0 | 1 | 0 | 1 | |
| | | | | | | | 1,0 | 1 | 0 | 0 | 1 | |
| | | | | | | | 1,1 | 1 | 1 | 1 | 1 | |
| | | | | $\|eg\rangle$ | 1 | 0 | 0,0 | 1 | 1 | 1 | 1 | |
| | | | | | | | 0,1 | 1 | 0 | 0 | 1 | |
| | | | | | | | 1,0 | 0 | 1 | 0 | 1 | |
| | | | | | | | 1,1 | 0 | 0 | 1 | 1 | |
| | | $\|ee\rangle$ | 0 | $\|gg\rangle$ | 0 | 0 | 0,0 | 0 | 1 | 1 | 1 | |
| | | | | | | | 0,1 | 0 | 0 | 0 | 1 | |
| | | | | | | | 1,0 | 1 | 1 | 0 | 1 | |
| | | | | | | | 1,1 | 1 | 0 | 1 | 1 | |
| | | | | $\|ee\rangle$ | 1 | 1 | 0,0 | 1 | 0 | 1 | 1 | |
| | | | | | | | 0,1 | 1 | 1 | 0 | 1 | |
| | | | | | | | 1,0 | 0 | 0 | 0 | 1 | |
| | | | | | | | 1,1 | 0 | 1 | 1 | 1 | |
| 1 | 0 | $\|gg\rangle$ | 0 | $\|gg\rangle$ | 0 | 0 | 0,0 | 1 | 0 | 1 | 1 | |
| | | | | | | | 0,1 | 1 | 1 | 0 | 1 | |
| | | | | | | | 1,0 | 0 | 0 | 0 | 1 | |
| | | | | | | | 1,1 | 0 | 1 | 1 | 1 | |
| | | | | $\|ee\rangle$ | 1 | 1 | 0,0 | 0 | 1 | 1 | 1 | |
| | | | | | | | 0,1 | 0 | 0 | 0 | 1 | |
| | | | | | | | 1,0 | 1 | 1 | 0 | 1 | |
| | | | | | | | 1,1 | 1 | 0 | 1 | 1 | |
| | | $\|ge\rangle$ | 1 | $\|ge\rangle$ | 0 | 1 | 0,0 | 1 | 1 | 1 | 1 | |
| | | | | | | | 0,1 | 1 | 0 | 0 | 1 | |
| | | | | | | | 1,0 | 0 | 1 | 0 | 1 | |
| | | | | | | | 1,1 | 0 | 0 | 1 | 1 | |
| | | | | $\|eg\rangle$ | 1 | 0 | 0,0 | 0 | 0 | 1 | 1 | |
| | | | | | | | 0,1 | 0 | 1 | 0 | 1 | |
| | | | | | | | 1,0 | 1 | 0 | 0 | 1 | |
| | | | | | | | 1,1 | 1 | 1 | 1 | 1 | |
| | | $\|eg\rangle$ | 1 | $\|ge\rangle$ | 0 | 1 | 0,0 | 1 | 1 | 1 | 1 | |
| | | | | | | | 0,1 | 1 | 0 | 0 | 1 | |
| | | | | | | | 1,0 | 0 | 1 | 0 | 1 | |
| | | | | | | | 1,1 | 0 | 0 | 1 | 1 | |
| | | | | $\|eg\rangle$ | 1 | 0 | 0,0 | 0 | 0 | 1 | 1 | |
| | | | | | | | 0,1 | 0 | 1 | 0 | 1 | |
| | | | | | | | 1,0 | 1 | 0 | 0 | 1 | |
| | | | | | | | 1,1 | 1 | 1 | 1 | 1 | |
| | | $\|ee\rangle$ | 0 | $\|gg\rangle$ | 0 | 0 | 0,0 | 1 | 0 | 1 | 1 | |
| | | | | | | | 0,1 | 1 | 1 | 0 | 1 | |
| | | | | | | | 1,0 | 0 | 0 | 0 | 1 | |
| | | | | | | | 1,1 | 0 | 1 | 1 | 1 | |
| | | | | $\|ee\rangle$ | 1 | 1 | 0,0 | 0 | 1 | 1 | 1 | |

| | | | | | | | | | | | |
|---|---|---|---|---|---|---|---|---|---|---|---|
| | | | | | | | 0,1 | 0 | 0 | 0 | 1 |
| | | | | | | | 1,0 | 1 | 1 | 0 | 1 |
| | | | | | | | 1,1 | 1 | 0 | 1 | 1 |
| 1 | 1 | $\|gg\rangle$ | 0 | $\|gg\rangle$ | 0 | 0 | 0,0 | 1 | 1 | 0 | 0 |
| | | | | | | | 0,1 | 1 | 0 | 1 | 0 |
| | | | | | | | 1,0 | 0 | 1 | 1 | 0 |
| | | | | | | | 1,1 | 0 | 0 | 0 | 0 |
| | | | | $\|ee\rangle$ | 1 | 1 | 0,0 | 0 | 0 | 0 | 0 |
| | | | | | | | 0,1 | 0 | 1 | 1 | 0 |
| | | | | | | | 1,0 | 1 | 0 | 1 | 0 |
| | | | | | | | 1,1 | 1 | 1 | 0 | 0 |
| | | $\|ge\rangle$ | 1 | $\|ge\rangle$ | 0 | 1 | 0,0 | 1 | 0 | 0 | 0 |
| | | | | | | | 0,1 | 1 | 1 | 1 | 0 |
| | | | | | | | 1,0 | 0 | 0 | 1 | 0 |
| | | | | | | | 1,1 | 0 | 1 | 0 | 0 |
| | | | | $\|eg\rangle$ | 1 | 0 | 0,0 | 0 | 1 | 0 | 0 |
| | | | | | | | 0,1 | 0 | 0 | 1 | 0 |
| | | | | | | | 1,0 | 1 | 1 | 1 | 0 |
| | | | | | | | 1,1 | 1 | 0 | 0 | 0 |
| | | $\|eg\rangle$ | 1 | $\|ge\rangle$ | 0 | 1 | 0,0 | 1 | 0 | 0 | 0 |
| | | | | | | | 0,1 | 1 | 1 | 1 | 0 |
| | | | | | | | 1,0 | 0 | 0 | 1 | 0 |
| | | | | | | | 1,1 | 0 | 1 | 0 | 0 |
| | | | | $\|eg\rangle$ | 1 | 0 | 0,0 | 0 | 1 | 0 | 0 |
| | | | | | | | 0,1 | 0 | 0 | 1 | 0 |
| | | | | | | | 1,0 | 1 | 1 | 1 | 0 |
| | | | | | | | 1,1 | 1 | 0 | 0 | 0 |
| | | $\|ee\rangle$ | 0 | $\|gg\rangle$ | 0 | 0 | 0,0 | 1 | 1 | 0 | 0 |
| | | | | | | | 0,1 | 1 | 0 | 1 | 0 |
| | | | | | | | 1,0 | 0 | 1 | 1 | 0 |
| | | | | | | | 1,1 | 0 | 0 | 0 | 0 |
| | | | | $\|ee\rangle$ | 1 | 1 | 0,0 | 0 | 0 | 0 | 0 |
| | | | | | | | 0,1 | 0 | 1 | 1 | 0 |
| | | | | | | | 1,0 | 1 | 0 | 1 | 0 |
| | | | | | | | 1,1 | 1 | 1 | 0 | 0 |

Table 2  Comparison between our protocol and some previous QPC protocols

| | Ref.[38] | Ref.[41] | Ref.[46] | Ref.[52] | Ref.[53] | Ref.[58] | Ref.[60] | Our protocol |
|---|---|---|---|---|---|---|---|---|
| Initial quantum resource | Bell entangled states | Two-particle product state | Bell entangled states | Four-particle W entangled states and Bell entangled states | GHZ entangled states | Three-particle W entangled states | Four-particle cluster entangled states | Two-atom product state |
| Quantum measurement (except that for security check) | Bell state measurements | single-particle measurements | single-particle measurements | Bell state measurements | single-particle measurements | single-particle measurements | single-particle measurements | Single-atom measurements |
| Use of QKD method | No (Use of Hash function instead) | Yes | No | Yes | No | Yes | No | Yes |
| Use of unitary operations | Yes | No | No | No | Yes | No | Yes | No |
| Use of entanglement swapping | No | No | No | Yes | No | No | No | No |
| TP's knowledge about the comparison result | No | Yes | Yes | Yes | Yes | Yes (with the probability of $1-\left(\frac{1}{2}\right)^{L-1}$) | Yes | Yes (with the probability of $1-\left(\frac{1}{2}\right)^{L-1}$) |
| Bit number compared each round | $L$ | 1 | 1 | $L$ | 1 | 1 | $L$ | 1 |
| Comparison times $n$ | 1 | $L$ | $L$ | 1 | $L$ | $L$ | 1 | $L$ |
| Qubit efficiency $\eta_e$ | 25% | 50% | 50% | 33% | 33% | 33% | 25% | 50% |

1